\documentclass[aps,preprintnumbers,amsmath,amssymb,superscriptaddress,nofootinbib]{revtex4-2}
\def\be{\begin{eqnarray}}
\def\ee{\end{eqnarray}}
\usepackage[dvipdfmx]{graphicx}
\usepackage{graphicx}
\usepackage{amsfonts,bm}

\usepackage{amsmath,amssymb,amsfonts,amssymb,bm,mathrsfs}

\usepackage{tikz}
\def\p{\partial}

\newcommand{\Exp}[1]{\left\langle~#1~\right\rangle}

\def\ben{\begin{equation}}
\def\een{\end{equation}}
\def\bena{\begin{eqnarray}}
\def\eena{\end{eqnarray}}

\def\mS{{\mathbb S}}

\begin{document}
\title{The averaged null energy condition on holographic evaporating black holes}

\author{Akihiro {\sc Ishibashi}}\email[]{akihiro@phys.kindai.ac.jp}
\affiliation{%
{\it Department of Physics and Research Institute for Science and Technology, Kindai University, Higashi-Osaka 577-8502, JAPAN
}}

\author{Kengo {\sc Maeda}}\email[]{maeda302@sic.shibaura-it.ac.jp}
\affiliation{%
{\it Faculty of Engineering,
Shibaura Institute of Technology, Saitama 330-8570, JAPAN}}

\begin{abstract}
We examine the averaged null energy condition~(ANEC) for strongly coupled fields, along the event horizon of an evaporating black hole 
by using the AdS/CFT duality. First, we consider a holographic model of a $3$-dimensional evaporating black hole with a perturbed 4-dimensional black droplet geometry 
as the bulk dual, and investigate how negative energy flux going into the boundary black hole horizon appears. 
We show that the ingoing negative energy flux always appears at the boundary black hole horizon when the horizon area decreases. 
Second, we test the ANEC in a holographic model whose boundary geometry is a 4-dimensional asymptotically flat spacetime, 
describing the formation and subsequent evaporation of a spherically symmetric black hole.  
By applying the ``bulk-no-shortcut principle'', we show that the ANEC is always satisfied when the local null energy 
is averaged with a weight function along the incomplete null geodesic on the event horizon from beginning of the formation 
to the final instant of the black hole evaporation. Our results indicate that the total ingoing negative energy flux is compensated by a large 
amount of positive energy flux in the early stage of the black hole formation. 
\end{abstract}
\maketitle

\section{Introduction}
Black hole evaporation is one of the most mysterious phenomena in quantum gravity. Hawking's discovery~\cite{Hawking1975} 
that a black hole evaporates by emitting thermal radiation is based on quantum field theory in a fixed curved background. 
There have been a number of calculations of the vacuum expectation value of the stress-energy tensor for quantum fields 
on various black hole backgrounds. 
In particular, negative energy flux going into the black hole horizon was found in a two-dimensional model 
of gravitational collapse for the Unruh vacuum state~\cite{DaviesFullingUnruh1976, Unruh1976}. 
This effect is consistent with the thermal radiation from the black hole 
observed at spatial infinity, according to the energy conservation. Because of the technical difficulties, however, most of the 
calculations are restricted to free field theories with no interaction. 
   
For interacting fields such as a strongly coupled field, the AdS/CFT correspondence~\cite{Maldacena1998} provides a useful 
tool to investigate the Hawking radiation on a fixed black hole background. 
So far, several models of AdS bulk solutions associated with 
boundary black holes have been numerically constructed~\cite{FiguerasLuciettiWiseman2011, SantosWay2012, FischettiSantos2013, FT2013, Mefford2017}. 
The black droplet is such an AdS bulk black hole solution in which the boundary black hole horizon is disconnected from the bulk horizons~\cite{HubenyMarolfRangamani2010}. When there are no non-extremal bulk horizons isolated from the AdS boundary, 
the black droplet solution is regarded as the gravity dual to the Unruh vacuum state in the boundary field theory, as the 
negative energy induced by a vacuum polarisation effect decays to zero at the spatial infinity~\cite{FiguerasLuciettiWiseman2011}. 
 
One peculiar behavior in strongly coupled fields in Unruh vacuum state is that there is no negative energy flux 
through the horizon, except two dimensional conformal field theory~\cite{FischettiMarolf2012}~\footnote{On the non-equilibrium 
thermal state, heat flow between two boundary black holes occurs for a non-stationary four-dimensional bulk solution~\cite{FischettiMarolfSantos2013}.}.    
This would be explained by the formation of quantum ``hair'' around the black hole~\cite{FiguerasLuciettiWiseman2011}. 
Due to its strong attractive self-interaction, the quantum field tends to collapse into the black hole, and it is 
balanced to the radiation pressure from the black hole. 
Although this picture is quite interesting, we need further investigations on a more general non-stationary black droplet solution 
containing a time-dependent boundary black hole, since most of the calculations so far made are based on the assumption 
that the AdS bulk spacetime is time-independent. 

In this paper, we consider evaporating black holes constructed on time-dependent boundaries of some asymptotically AdS bulk spacetime, 
for which, as shown in Sec. \ref{sec:2}, the null-null component of the boundary Ricci tensor $R_{\mu\nu}l^\mu l^\nu$ is found to be negative, 
where $l^\mu$ is the tangent vector to a null geodesic generator of the boundary black hole horizon. 
This does not immediately imply that the null-null component of the vacuum expectation value of the boundary stress-energy tensor 
$\Exp{T_{\mu\nu}}l^\mu l^\nu$ is necessarily negative because the AdS/CFT correspondence itself does not provide 
any dynamical equations of motion such as $R_{\mu\nu}l^\mu l^\nu=8\pi G\Exp{T_{\mu\nu}}l^\mu l^\nu$ on the boundary. 
However, we can still expect that $\Exp{T_{\mu\nu}}l^\mu l^\nu$ is negative, or negative energy flux across the horizon appears when 
the horizon area decreases from the perspective of the conjectured generalized second law~(GSL) as follows. 
The GSL states that the sum of the gravitational and 
the matter entropy outside the black hole cannot decrease even during the evaporation process~\cite{Bekenstein1973, Hawking1975}. 
The GSL is widely believed to be true and it was established for super-renormalizable quantum field 
theories~\cite{Wall2012}~(for a holographic setting, see also~\cite{BuntingFuMarolf2016}). If the GSL is always satisfied for any 
quantum field theories, ingoing negative energy flux should appear on the horizon when the area decreases, in accord with 
the energy conservation law. 
In this paper, we consider two such time-dependent holographic models; one is a $3$-dimensional circular symmetric boundary 
black hole and the other a $4$-dimensional spherically symmetric boundary black hole. In the $3$-dimensional case, we will explicitly 
show that ingoing negative energy flux always appears on the horizon when the area decreases, in accord with 
the GSL picture. 

For the $3$-dimensional time-dependent model, we start from the $4$-dimensional static black droplet 
solution~\cite{HubenyMarolfRangamani2010_Ver2} as our background bulk geometry, whose boundary metric is conformal to the $3$-dimensional Banados-Teitelboim-Zanelli~(BTZ) black hole~\cite{BTZblackhole}. Then, we consider metric perturbations on this background 
so that the perturbed geometry describes a regular bubble in the bulk and an evaporating black hole on the boundary. 
We are concerned with what happens when we dynamically perturb the static BTZ black hole with Unruh vacuum. 
As expected from the above observation, we show that a negative energy flux going into the boundary black hole horizon always appears 
when the horizon area decreases under the adiabatic approximation. From the perspective of the AdS/CFT correspondence, 
the stress-energy tensor on the boundary field theory depends on both of the boundary geometry and the boundary condition deep inside the bulk. 
Since a regularity condition is imposed on the bubble radius in our model as a natural boundary condition, the stress-energy tensor is only determined 
by the boundary geometry. So, our results indicate that the negative energy flux reflects the negative value of 
$R_{\mu\nu}l^\mu l^\nu$ on the boundary black hole horizon. 


Another key question regarding the negative energy flux is whether there is a lower bound for the flux. 
Presumably, the absolute value of the total negative energy flux would be bounded from above by the initial black hole mass 
by energy conservation law. 
To evaluate the total amount of the flux, we consider a holographic model in which the AdS boundary metric is conformal to a 
$4$-dimensional asymptotically flat spherically symmetric metric which describes an evaporating black hole formed by a gravitational collapse. 
The null geodesic generator of the event horizon begins at a regular spacetime point and ends at a zero mass naked singularity,  
where the evaporation is completed. The behavior of the null geodesic congruence is very similar to the one on the 
spatially compact $S^3$ universe in which the null geodesic congruence expands from a point on the south pole and 
shrinks again to a point on the north pole. In such spatially compact universes, it was shown that 
the averaged null energy condition~(ANEC) with a weight function is satisfied along the null geodesic under 
the ``no-bulk-shortcut condition'', stating that no bulk causal curve can travel faster than the boundary 
achronal null geodesics~\cite{IizukaIshibashiMaeda2020, IizukaIshibashiMaeda2020_Ver2}. 
When the null energy condition is satisfied in the bulk spacetime, the no-bulk-shortcut condition is satisfied under the 
assumptions that there are no pathological behavior such as a naked singularity formation in the bulk and the 
boundary~\cite{GaoWald2000}. 
By applying the no-bulk-shortcut condition to the holographic model of the evaporating 
black hole, we show that the ANEC is satisfied.   

The paper is organized as follows. In Section \ref{sec:2}, we consider the $3$-dimensional evaporating black hole on the boundary of the perturbed 
$4$-dimensional static black droplet solution and derive the negative energy flux going into the boundary black hole horizon. 
In Section \ref{sec:3}, we derive the ANEC with a weight function in a background of a $4$-dimensional evaporating black hole 
in the context of the AdS/CFT correspondence. Section \ref{sec:4} is devoted to summary and discussions.

\section{$3$-dimensional evaporating black hole and negative energy flux}
\label{sec:2}
In this section, we construct a $4$-dimensional asymptotically locally AdS spacetime with a $3$-dimensional 
time-dependent black hole on the AdS boundary, describing the evaporation of a black hole. 
We assume that the evaporating process of our boundary black hole proceeds very slowly so that the boundary black hole can be viewed as 
almost static. Then, the corresponding 4-dimensional bulk dual should also be well described by an almost static or stationary black droplet 
geometry which has no non-extremal bulk horizon isolated from the boundary black hole horizon.  
We construct such a bulk geometry by perturbing the exact background solution of a static or stationary black droplet.

\subsection{A holographic model}
We consider an analytic four-dimensional black droplet solution of the four-dimensional vacuum Einstein 
equations with a negative cosmological constant~\cite{HubenyMarolfRangamani2010_Ver2}, in which 
there is a bubble of nothing deep inside the bulk, and the boundary metric is conformal to the BTZ black hole.  
The metric of the $4$-dimensional black droplet solution is written by  
\begin{align} 
\label{BTZ_string}
ds^2=\frac{d\rho^2}{f(\rho)}+\frac{\rho^2r_0^2}{r^2l^2}\left[-\frac{r^2-r_0^2}{l^2}dt^2+\frac{l^2dr^2}{r^2-r_0^2}
+L^2\frac{r^2f(\rho)}{\rho^2}d\varphi^2  \right],  \qquad 
f(\rho)=\frac{\rho^2}{L^2}+1 - \dfrac{\mu_0}{\rho} \,, 
\end{align}
where $L$ and $l$ denote the AdS curvature length in the bulk and that in the boundary, respectively, and where $\mu_0$ describes 
the mass parameter given by 
\be
&{} 
\mu_0 := \rho_0 \left( \dfrac{\rho_0^2}{L^2}+ 1 \right) \,.    
\ee 
Here, $\varphi$ is the periodic coordinate with period $2\pi$, and the circle along $\varphi$ smoothly caps off at the bulk radius, $\rho=\rho_0$ when the horizon radius $r_0$ of the boundary BTZ black hole is given by 
\be 
\label{def_r+}
r_0=\frac{2\rho_0lL}{3\rho_0^2+L^2} \,. 
\ee
In this case, a bubble appears at the radius, $\rho=\rho_0$ instead of the non-extremal bulk horizon.

Since the boundary metric of Eq.~(\ref{BTZ_string}) at infinity $\rho \rightarrow \infty$ is conformal to the static BTZ black hole, the solution describes a confining 
vacuum in the dual field theory in the BTZ black hole background.  
Following the procedure~\cite{HaroSkenderisSolodukhin2001}, the stress-energy tensor is calculated as 
\be 
& T_{tt}=-\dfrac{\mu_0 r_0^3 L\left(r^2-r_0^2\right)}{16\pi G_4 l^4r^3}, \quad T_{rr}=\dfrac{L \mu_0 r_0^3}{16\pi G_4 r^3 (r^2-r_0^2)} \,, \quad
T_{\varphi\varphi}=-\dfrac{\mu_0 Lr_0^3}{8\pi G_4 l^2 r} \,, 
\ee
where $G_4$ is the 4-dimensional gravitational constant. 
It is clear that there is no energy flux on the boundary black hole horizon, $r=r_0$, even though the energy density outside the BTZ black hole 
is negative. In the following subsections, we construct a time-dependent boundary evaporating black hole geometry by considering 
perturbations of the static droplet solution~(\ref{BTZ_string}).

\begin{figure}[htbp]
 \begin{center}
  \includegraphics[width=80mm]{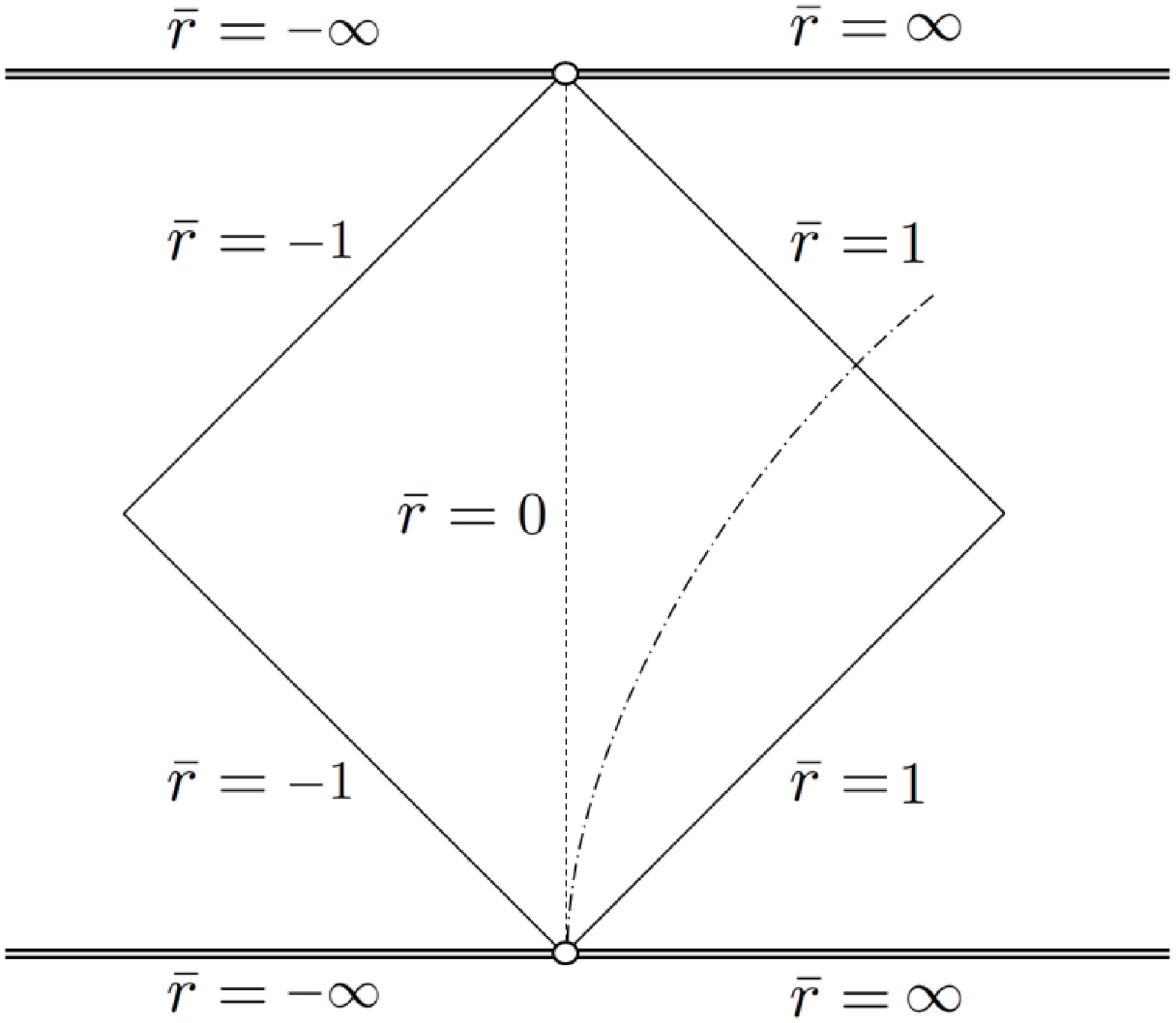}
 \end{center}
 \caption{The Penrose diagram of the two-dimensional de Sitter spacetime. The dot-dashed curve represents 
an imaginary surface of the matter of the gravitationally collapse.}
 \label{fig:one}
\end{figure}

\subsection{Metric perturbations in the 4-dimensional bulk}\label{subsec:3d:perturbation}

Under the coordinate transformation $(r,t,\varphi) \rightarrow (\bar{r}, \bar{t}, \bar{\varphi})$, 
\be 
\label{transform_tilde_r}
r=\frac{r_0}{\bar{r}}, \quad t=\frac{l^2\bar{t}}{r_0}, \quad \varphi=\frac{l}{Lr_0} \bar{\varphi}\, , 
\ee
the metric~(\ref{BTZ_string}) becomes the following warped product type metric that includes, as a part, the $2$-dimensional
de Sitter metric, 
\be 
\label{deSitter}
ds^2 &=f(\rho)d \bar{\varphi}^2+\dfrac{d\rho^2}{f(\rho)}+\rho^2 \left[-(1-\bar{r}^2)d\bar{t}^2+\dfrac{d\bar{r}^2}{1-\bar{r}^2}\right] \,.   
\ee
Note that the de Sitter horizon at $\bar{r}=1$ corresponds to the BTZ black hole horizon at $r=r_0$ on the boundary, as can be seen in~(\ref{BTZ_string}).

In order to consider perturbations on this geometry, it is more convenient to take the double wick rotation: 
\be  
\label{Wick_rotation}
 \bar{\varphi}=i \tau, \quad \bar{t}= i \phi, \quad \bar{r}=\cos\theta \,. 
\ee
Then the black droplet metric takes the standard form of the $4$-dimensional Schwarzschild AdS metric:
\be
\label{met:SchAdS}
ds^2 = g_{ab} (y)dy^2dy^b + \rho^2 \gamma_{ij}(z) dz^idz^j = - f(\rho) d\tau^2 + \dfrac{d\rho^2}{f(\rho)} + \rho^2 \left(d \theta^2 + \sin^2\theta d\phi^2 \right) \,,
\ee
where here and hereafter we use the coordinate notation $y^a=(\tau, \rho)$ and $z^i=(\theta, \phi)$. 
On this background we perform metric perturbations by applying the gauge-invariant formalism of~\cite{KodamaIshibashi2003}. 
For our purpose, it is sufficient to consider the scalar(polar)-type metric perturbations, which can be expanded in terms of 
the scalar harmonic functions $\mS{}_{\bf k}$ on the unit $2$-sphere, 
\be
\label{def:mS}
 \hat{D}^i\hat{D}_i \mS{}_{\bf k} + k^2 \mS{}_{\bf k} =0 \,, 
\ee 
where $\hat{D}_i$ denotes the derivative operator associated with the unit $2$-sphere metric $\gamma_{ij}dz^i dz^j= d \theta^2 + \sin^2\theta d\phi^2$.  
We can express the components $(\delta g_{ab}, \delta g_{ai}, \delta g_{ij})$ of the scalar-type metric perturbations as 
\be 
\label{pert_variable}
& \delta g_{ab}=\epsilon f_{ab {\bf k}}\mS_{\bf k}, \quad \delta g_{a j }=\epsilon \rho f_{a {\bf k}}\hat{D}_j \mS_{\bf k} \,, 
\quad  
\delta g_{ij}=2\epsilon \rho^2(H_{L {\bf k}} \gamma_{ij}\mS_{\bf k}+H_{T{\bf k}}\mS_{\bf (k)}{}_{ij}) \,, 
\ee
where $\epsilon$ is an infinitesimally small parameter and where $\mS_{\bf k}{}_{ij}$ is defined by 
\be
\label{def:Sij}
& \mS_{\bf (k)}{}_{ij}=\dfrac{1}{k^2}\hat{D}_i \hat{D}_j \mS_{\bf k}+\dfrac{1}{2}\gamma_{ij }\mS_{\bf k} \,. 
\ee
Hereafter we omit the mode index ${\bf k}$, for the notational simplicity. 
We introduce the vector field $X_a$ on the $2$-dimensional spacetime spanned by $y^a=(\tau, \rho)$ by   
\be
\label{vector_X} 
X_a=\dfrac{\rho}{k}\left(f_a+\dfrac{\rho}{k}D_aH_T\right) \,,  
\ee 
where $D_a$ is the derivative operator associated with the $2$-dimensional metric $g_{ab}$. 
We find the following combinations are gauge invariant;   
\be
\label{gauge-invariant}
& F=H_L+\dfrac{1}{2}H_T+\dfrac{1}{\rho}D^a\rho\, X_a \,, \quad 
F_{ab}=f_{ab}+D_aX_b+D_bX_a \,.  
\ee
For convenience, we introduce 
\be
& X :={F_\tau}^\tau-2F, \quad Y={F_\rho}^\rho-2F \,. 
\ee
Furthermore, we assume that $\p_\tau$ is a Killing vector on the perturbed spacetime. Thus, we consider 
the static perturbations in the background (\ref{met:SchAdS}). 
This, in turn, implies that we are concerned with dynamical but circular perturbations along $\varphi$ in the original black droplet background via the double wick rotation (\ref{Wick_rotation}) and the rescaling (\ref{transform_tilde_r}). In this case, $X$, $F$, and ${F_a}^b$ are determined by $Y$ as    
\be
\label{X_Y_F_relation} 
& X= Y + 2\dfrac{f}{f'}Y \,, \quad {F_\tau}^\tau=-{F_\rho}^\rho=\dfrac{f}{f'}Y', \quad F=-\dfrac{1}{2}\left( Y + \dfrac{f}{f'}Y' \right) \, , 
\ee
where the prime denotes the derivative with respect to $\rho$. 

For later convenience, we choose the gauge 
\be 
\label{gauge_rho}
f_\rho=f_{\rho\rho}=f_\tau=0 \,. 
\ee
Then, $X_\tau=0$ for the static perturbation. By Eqs.~(\ref{gauge-invariant}), (\ref{X_Y_F_relation}), and (\ref{gauge_rho}), 
$X_\rho$, $H_T$, and $H_L$ are determined by $Y$ as 
\be 
\label{X_rho_HT_HL}
& X_\rho=-\dfrac{1}{2\sqrt{f}} \displaystyle\int \sqrt{f}\,\dfrac{Y'}{f'} d\rho, \quad H_T=k^2\int \dfrac{X_\rho}{\rho^2}d\rho \,, 
\quad H_L=-\dfrac{1}{2}\left(Y+\dfrac{fY'}{f'}\right)-\dfrac{H_T}{2}-\dfrac{f}{\rho}X_\rho \,.
\ee
The master variable $Y$ satisfies  
\be 
\label{master_Y}
& Y''+\alpha Y'+\beta Y=0, 
\ee
where 
\be
\rho^2 f f'\alpha=4x-\dfrac{4\rho^2}{L^2}-2x^2+\dfrac{16}{L^2}x\rho^2+\dfrac{4\rho^4}{L^4} \,, \quad 
\rho^2 f \beta=-k^2+2 \,,
\ee
with $x:= {\mu_0}/{\rho}$. 

\subsection{The boundary conditions in the bulk}

Let us consider appropriate boundary conditions for our perturbations.  
We first consider boundary conditions from the bulk view point, which we need to impose at the bubble radius, $\rho=\rho_0$, and 
at the AdS infinity, $\rho=\infty$. The former condition comes from the regularity condition~\cite{MarsSenovilla1993}\footnote{
The right-hand side of (\ref{condi:reg}) comes from the fact that $\bar{\varphi}$ has the period $2 \pi Lr_0/l$ via the rescaling 
(\ref{transform_tilde_r}). If $\xi^2$ is chosen to be the norm squared of $\partial/\partial \varphi$, the r.h.s should be unit since 
the $\varphi$ circle has period $2\pi$. 
}, 
\be 
\label{condi:reg}
\lim_{\rho\to \rho_0}\dfrac{\nabla_M(\xi^2)\nabla^M(\xi^2)}{4\xi^2}= \left(\dfrac{l}{Lr_0} \right)^2 \,, 
\ee 
where $\xi^2$ denotes the norm squared of the circular Killing vector $\partial/\partial \bar{\varphi}$. Note that $\bar{\varphi}= i\tau$, and therefore 
that the r.h.s is equivalent to square of the surface gravity $\kappa= l/Lr_0$ of the Killing horizon with respect to $\partial/\partial \tau$ in the Schwarzschild-AdS black hole (\ref{met:SchAdS}). 
Up to $O(\epsilon)$, this regularity condition reduces to 
\be 
f_{\tau \tau }=\alpha_{ \tau \tau }(\rho-\rho_0)^2+\cdots. 
\ee 
This determines the boundary condition for $f_{\tau \tau}$ and it is derived from Eqs.~(\ref{gauge-invariant}) and (\ref{X_rho_HT_HL}) as 
\be
\label{f_TT}
 f_{\tau \tau}=F_{\tau \tau}+ff'X_\rho=-\dfrac{f^2}{f'}Y'-\dfrac{\sqrt{f}}{2}f'\int^\rho_{\rho_0} \sqrt{f}\dfrac{Y'}{f'}d\rho \,, 
\ee
which is equivalent to the condition 
\be 
\label{bc_X_rho}
X_\rho(\rho_0)=0 \,. 
\ee

Note that there is another integration constant which remains undetermined for $H_T$ in Eq.~(\ref{X_rho_HT_HL}), while      
the integration constant of $X_\rho$ is determined by the regularity condition, $X_\rho(\rho_0)=0$. 
To determine the integration constant for $H_T$, we impose the following condition at the AdS boundary: 
\be 
\label{H_T_bc_infinity}
\lim_{\rho\to \infty}\dfrac{H_T}{\rho^2}=0 .
\ee
This condition for $H_T$, together with our gauge condition~(\ref{gauge_rho}) and the coordinate transformation~(\ref{transform_tilde_r}), implies 
that our perturbations deform the boundary BTZ metric to the following form; 
\be 
ds^2=e^{A(t,r)}\left(-\dfrac{r^2-r_0^2}{l^2}dt^2+\dfrac{l^2dr^2}{r^2-r_0^2}\right)
+e^{B(t,r)}r^2d\varphi^2 \,.  
\ee   
Namely, the perturbed boundary black hole is time-dependent and circularly symmetric as intended, 
but the coordinate location of the event horizon is fixed at $r=r_0$ under our gauge. 

\subsection{Boundary conditions on the boundary}\label{subsec:3d:bc:boundary}
Let us turn to boundary conditions from the boundary view point, which are to be imposed at the horizon of 
the 3-dimensional boundary black hole. 
For this purpose, we take the double wick rotation (\ref{Wick_rotation}) again and go back to the background geometry (\ref{deSitter}). 
By doing so, we can now view that our metric perturbations considered in the previous subsection~\ref{subsec:3d:perturbation} are expanded 
by---instead of the harmonic functions on the $2$-sphere defined in (\ref{def:mS})---the mode functions $\mS_{\bf k}$ on the $2$-dimensional de Sitter spacetime spanned by $(\bar{t},\bar{r})$, which now satisfy the wave equation below,    
\be 
\label{harmonic_function}
   \left[ 
           - \dfrac{\partial^2}{\partial {\bar{t}}^2}  + (1-\bar{r}^2)\left\{ \dfrac{\partial}{\partial \bar{r}}(1-\bar{r}^2)\dfrac{\partial}{\partial \bar{r}} \right\}  
            + (1-\bar{r}^2)k^2 
   \right] \mS_{\bf k} = 0 \,.    
\ee   
We set $\mS_{\bf k}=e^{-i\omega \bar{t}}S_{\bf k}(\bar{r})$ so that $S_{\bf k}(\bar{r})$ satisfies 
\be
 \left[(1-\bar{r}^2)^2\dfrac{d^2 }{d\bar{r}^2}-2\bar{r}(1-\bar{r}^2) \dfrac{d}{d \bar{r}} + k^2 (1-\bar{r}^2) + \omega^2  \right] S_{\bf k}=0 \,,
\ee
and find that the general solution of $S_{\bf k}(\bar{r})$ is represented by the associated 
Legendre functions, $P^m_n$ and $Q^m_n$ as 
\be
\label{general_S}
S_{\bf k}=a\,P^m_n(\bar{r})+b\,Q^m_n(\bar{r}), \quad m:=i\omega, \,\, n:=\frac{1}{2}(-1+\sqrt{1+4k^2}) \,, 
\ee
where 
\be 
\label{associated_Legendre}
& P^m_n(x)=\dfrac{e^{m\pi i}}{\Gamma(1-m)}\left(\dfrac{1+x}{1-x} \right)^{m/2}F \left(-n, n+1, 1-m; \dfrac{1-x}{2}\right), 
\nonumber \\
& Q^m_n(x)=\dfrac{\pi}{2\sin m\pi}\left[\cos m\pi\, P^m_n(x)-\dfrac{\Gamma(n+m+1)}{\Gamma(n-m+1)}P^{-m}_n(x)   \right]. 
\ee
Here, note that we assume that $|m|$ is very small, since we consider the case that the time evolution of the 
bulk geometry is very slow. 

Since we are interested in the future horizon, $\bar{r}=1$ at the right-hand side (r.h.s.) in Fig.~\ref{fig:one},  
the ingoing boundary condition should be imposed as a regularity condition. In terms of a retarded null coordinate $u$, 
$S_{\bf k}$ behaves near the horizon as  
\be
\label{retarded_null_future} 
\mS_{\bf k}\sim e^{-i\omega u}, \quad u:=\bar{t}-\int \dfrac{d\bar{r}}{1-\bar{r}^2}\simeq \bar{t}+\dfrac{1}{2}\ln (1-\bar{r}). 
\ee 
The ingoing mode comes from the left past horizon, $\bar{r}=-1$ in Fig.~\ref{fig:one}, as an outgoing wave. So, we shall 
impose that $S_{\bf k}$ behaves near the past horizon, $\bar{r}=-1$ at the left-hand side (l.h.s.) of  Fig.~\ref{fig:one} as  
\be
\label{retarded_null_past} 
\mS_{\bf k}\sim e^{-i\omega u}, \quad u:=\bar{t}-\int \dfrac{d\bar{r}}{1-\bar{r}^2}\simeq \bar{t}-\dfrac{1}{2}\ln (1+\bar{r}). 
\ee 
From the boundary condition~(\ref{retarded_null_future}) and the form of the associated Legendre function~(\ref{general_S}),  
$S_{\bf k}\sim P^{m}_n(x)$ near the future horizon, $\bar{r}=1$, and then, we obtain 
\be 
\label{coeff_b}
b=0. 
\ee
Using the linear transformation formulae for hypergeometric functions,  $S_{\bf k}$ behaves near the past horizon, $\bar{r}=-1$ as 
\be 
\label{linear_transformation}
& S_{\bf k}\sim
P^{m}_n(\bar{r})\simeq e^{m\pi i}\left(\dfrac{1+\bar{r}}{2}\right)^\frac{m}{2}\Biggl[ \dfrac{\Gamma(m)}{\Gamma(-n)\Gamma(n+1)}
\left(\dfrac{1+\bar{r}}{2}\right)^{-m}F\left(1+n-m, -n-m, 1-m; \dfrac{1+\bar{r}}{2}\right) \nonumber \\
&+\dfrac{\Gamma(-m)}{\Gamma(1-m+n)\Gamma(-n-m)}
F\left(-n, n+1, m+1; \dfrac{1+\bar{r}}{2}\right)\Biggr] \,. 
\ee
By imposing the boundary condition~(\ref{retarded_null_past}), we find that the first term should be zero, and thus, 
\be 
\label{integer}
n=N,  \quad k^2=N(N+1)
\ee 
for an arbitrary positive integer $N$. Note that here, we do not consider a particular class of the mode function, $k=0$ in which 
$H_T=0$ in Eq.~(\ref{pert_variable}).   
In Sec. \ref{sec:3} we will consider a $4$-dimensional time-dependent boundary geometry that describes a black hole 
formation and evaporation via Hawking radiation, in which the energy flux comes in from past null infinity. 
Our boundary conditions~(\ref{coeff_b}) and (\ref{integer}) may be viewed as boundary conditions that correspond to such a black hole 
formation and evaporation induced via some incoming flux from past infinity.

\subsection{Negative energy flux on the evaporating black hole}
\label{sec:5}
In this section we derive the stress-energy tensor of the boundary field theory by taking the Fefferman-Graham coordinate system 
and using the AdS/CFT dictionary~\cite{HaroSkenderisSolodukhin2001}, and then we evaluate it on the horizon. 
Thanks to the gauge choice~(\ref{gauge_rho}),  the Fefferman-Graham coordinates system 
\begin{align}
\label{FG_coordinate}
& g_{MN} dx^M dx^N=\frac{L^2}{z^2} \left( dz^2+g_{mn}(x,z)dx^m dx^n \right), \nonumber \\
& g_{mn}(z,x)=g_{(0)mn}(x)+z^2g_{(2)mn}(x)+\cdots 
\end{align}
is derived from
\be 
\label{rho_z_relation}
\rho(z)=\dfrac{L}{z}\left(1-\dfrac{1}{4}z^2+\dfrac{\mu }{6L}z^3+\cdots   \right) \,.  
\ee 
The stress-energy tensor $T_{mn}$ can be read off from the coefficient $g_{(3)mn}(x)$ as 
\be
\label{3-dim_stress_energy} 
T_{mn}=\dfrac{3 L^2}{16\pi G_4}g_{(3)mn}, 
\ee
where $G_4$ is the bulk gravitational constant. To derive the coefficient $g_{(3)mn}(x)$, let us expand 
the master variable $Y$ near the infinity as 
\be 
\label{expansion_Y}
Y=a_0+\dfrac{a_1}{\rho}+\dfrac{a_2}{\rho^2}+\dfrac{a_3}{\rho^3}+\cdots. 
\ee
From Eq.~(\ref{master_Y}), each coefficient $a_n~(n\ge 2)$ is determined by $a_0$ and $a_1$ as 
\be 
\label{higher_coeff_a_i}
& a_2=\dfrac{1}{2}a_0(k^2-2)L^2, \qquad a_3=\dfrac{1}{6}a_1(k^2-6)L^2, \nonumber \\ 
& a_4=\dfrac{1}{24}\{a_0(24-14k^2+k^4)L^4+18a_1\rho_0(L^2+\rho_0^2)   \} \,, \cdots. 
\ee
Under the boundary conditions (\ref{bc_X_rho}) and (\ref{H_T_bc_infinity}), $H_T$, $H_L$, and $f_{\bar{\varphi} \bar{\varphi}}$ are expanded as 
a series of $1/\rho$ from Eqs.~(\ref{X_rho_HT_HL}) and (\ref{f_TT}) as 
\be 
\label{asymp_H_T}
H_T=\dfrac{c_1k^2L}{4\rho^2}+\dfrac{a_1k^2L^2}{12\rho^3}+\cdots \,, 
\ee
\be
\label{asymp_H_L}
H_L &=&\dfrac{c_1-a_0L}{2L}+\dfrac{L(k^2-2)(a_0L-c_1)}{8\rho^2}+\dfrac{a_1(k^2-2)L^3-6c_1\rho_0(L^2+\rho_0^2)}{24L\rho^3}+\cdots \,, 
\nonumber \\
f_{\bar{\varphi} \bar{\varphi}} &=& \dfrac{c_1\rho^2}{L^3} + \dfrac{1}{4}\left\{a_0(k^2-2)-\dfrac{2c_1}{L} \right\} - \dfrac{a_1(k^2-2)}{6\rho}+\cdots \,, 
\ee
where the coefficient $c_1$ is defined as 
\be
\label{def:c_1} 
c_1:=\int^\infty_{\rho_0} \sqrt{f}\,\dfrac{Y'}{f'} d\rho \,. 
\ee

Substituting (\ref{asymp_H_L}) into (\ref{pert_variable}) and expanding the metric as a series in $z$ with the help of 
Eq.~(\ref{rho_z_relation}), we obtain the boundary metric, 
\be 
\label{deformed_bmetric}
g_{(0)mn}dx^m dx^n
 &=&\dfrac{r_0^2}{l^2r^2}\left[r^2\left(1+\epsilon \dfrac{c_1\mS_{\bf k}}{L}\right)d\varphi^2
+\left(1+\epsilon \dfrac{c_1-a_0L}{L}\mS_{\bf k} \right)\left(2dvdr-\dfrac{r^2-r_0^2}{l^2}dv^2   \right)\right] \nonumber \\
 &=:&\left(\dfrac{r_0}{r}\right)^2ds_{\rm dBTZ}^2 \,, 
 \ee
where 
\be
v :=t+\int \dfrac{l^2}{r^2-r_0^2}dr \,.
\ee 
Note that the stress-energy tensor for the boundary metric $ds_{\rm dBTZ}^2$---which stands for the deformed BTZ metric---cannot be proportional 
to the coefficient $g_{(3)mn}$, as $g_{(0)mn}$ is different from the metric $ds_{\rm dBTZ}^2$ by a conformal 
factor, $(r_0/r)^2$. In general, for the conformal transformation
\be 
\hat{g}_{mn}dx^m dx^n=\Omega^2g_{mn}dx^m dx^n \,, 
\ee
the stress-energy tensor transforms as 
\be 
\hat{T}_{mn}=\frac{1}{\Omega}T_{mn}  \,,
\ee
for the 3-dimensional boundary spacetime. Since we are interested in the stress-energy tensor on the horizon, $r=r_0$, 
the conformal factor is equal to $1$, and then, the null-null component of the stress energy tensor for the metric 
$ds_{\rm dBTZ}^2$ can be read off from the formula~(\ref{3-dim_stress_energy}) as
\be 
\label{null_null_component}
T_{vv} (v, r_0)=\dfrac{\epsilon a_1L}{32\pi G_4}\left( \partial_v^2 {\mS_{\bf k}} - \dfrac{r_0}{l^2} \partial_v{\mS_{\bf k}} \right) \,. 
\ee
The Ricci tensor for the deformed BTZ metric $ds_{\rm dBTZ}^2$ on the horizon is also calculated as 
\be
\label{Ricci_null_null} 
R_{vv} (v, r_0)=\epsilon \dfrac{c_1}{2L}\left(\dfrac{r_0}{l^2} \partial_v{\mS_{\bf k}} - \partial_v^2{\mS_{\bf k}} \right) \,.  
\ee
So, the ratio between $T_{vv} (v, r_0)$ and $R_{vv}  (v, r_0)$ is determined by the dimensionless 
coefficient $\zeta$,  
\be 
\label{coefficient_xi}
\zeta:= \frac{-a_1}{c_1} \,. 
\ee
As shown in the proposition below, this coefficient $\zeta$ is always positive, except the trivial case, $a_1=c_1=0$. \\
\\

\medskip 

\noindent 
{\it Proposition} \\
For the mode functions $\mS_{\bf k}$ satisfying Eq.~(\ref{harmonic_function}) with (\ref{integer}), $\zeta>0$ for all integers 
$N>1$, and $a_1=c_1=0$ for $N=1$. \\
{\it proof}) \\
In terms of the new variable, $u=\rho_0/\rho$, the master Eq.~(\ref{master_Y}) can be rewritten as 
\be
\label{Eq:Y_u}
&{}& (1-u)\{\hat{\rho}_0^2(1+u)+(1+\hat{\rho}_0^2)u^2 \}\{2\hat{\rho}_0^2+(1+\hat{\rho}_0^2)u^3   \}\ddot{Y}
\nonumber \\
&{}& \qquad 
-2u\{-4\hat{\rho}_0^2+9\hat{\rho}_0^2(1+\hat{\rho}_0^2)u+(1+\hat{\rho}_0^2)u^3   \}\dot{Y}
+(2-k^2)\{2\hat{\rho}_0^2+(1+\hat{\rho}_0^2)u^3   \}Y=0 \,, 
\nonumber \\
&{}& \dot{Y}(1)=\dfrac{1-\dfrac{k^2}{2}}{1+3\hat{\rho}_0^2}Y(1) \,, 
\ee
where $\hat{\rho}_0=\rho_0/L$, and the dot denotes the derivative with respect to $u$. The second equation 
comes from the regularity on the bubble radius, $\rho=\rho_0$. Since the differential equation is linear,  
we can set $Y(1)=1$ without loss of generality. First, consider the case $N>1$. Then, $k^2>2$ and hence $\dot{Y}(1)<0$.  
Now suppose $Y$ would have a maximum at $u=u_m~(0\le u_m \le 1)$. In this case, $Y(u_m)>1$, $\dot{Y}(u_m)=0$, and 
$\ddot{Y}(u_m)\leq 0$. This is impossible from Eq.~(\ref{Eq:Y_u}). So, $\dot{Y}<0$ for $0\le u\le 1$. In terms of $u$, 
the coefficient $c_1$ in Eq.~(\ref{def:c_1}) is rewritten by 
\be 
\label{c_1_u}
\frac{c_1}{L}=-\int^1_0 \dfrac{\sqrt{g(u)}\,\dot{Y}}{2+\left(1+\dfrac{1}{\hat{\rho}_0^2} \right)u^3}du \,, \qquad 
g(u):=1+\dfrac{u^2}{\hat{\rho}_0^2}-\left(1+\dfrac{1}{ \hat{\rho}_0^2} \right)u^3 \,. 
\ee
From this, $g(u)\ge 0$ and $c_1>0$, while $a_1< 0$ from the expansion~(\ref{expansion_Y}). Thus, we have shown $\zeta>0$. 
As for the $N=1$ case, the $Y$ term in Eq.~(\ref{Eq:Y_u}) vanishes, as $k^2=2$. This is analytically solved and the only regular solution is $Y=1$. In this case, 
$a_1=c_1=0$.  

\hfill $\Box$.   

\medskip 

Now let us consider the time dependence of our boundary black hole. By Eq.~(\ref{deformed_bmetric}), 
the area on the horizon is given by 
\be 
{\cal A}=2\pi r_0\sqrt{1+\epsilon \dfrac{c_1\mS_{\bf k}}{L}}\sim r_0\left(1+\dfrac{\epsilon c_1}{2L} \mS_{\bf k} \right) \,, 
\ee
where $\mS_{\bf k}$ is evaluated at $u=1$. 
By setting $0<m=i\omega\ll 1$ and choosing the sign of $\epsilon$ so that $\epsilon c_1\mS_{\bf k}>0$, 
the area decreases very slowly, thus describing the evaporating process. 
In this case, $\partial_v^2 \mS_{{\bf k}}$ is negligible compared with $\partial_v \mS_{{\bf k}}$, and hence,  
$R_{vv}(v, r_0)$ in Eq.~(\ref{Ricci_null_null}) is negative on the horizon. 
The above proposition means that the null-null component of the stress-energy tensor is always negative 
during the evaporating phase. By the energy conservation law, this implies that the energy outside the evaporating 
black hole increases. This picture agrees with the Hawking radiation, and thus, the entropy outside the horizon would 
increase, supporting the GSL conjecture.  

At first glance, the existence of the negative energy flux would violate the averaged null energy condition~(ANEC) on the 
horizon. In the next section, we examine the ANEC in a background of $4$-dimensional evaporating black hole 
in the context of the AdS/CFT correspondence.

\section{$4$-dimensional evaporating black hole and ANEC}
\label{sec:3}
\begin{figure}[htbp]
 \begin{center}
  \includegraphics[width=70mm]{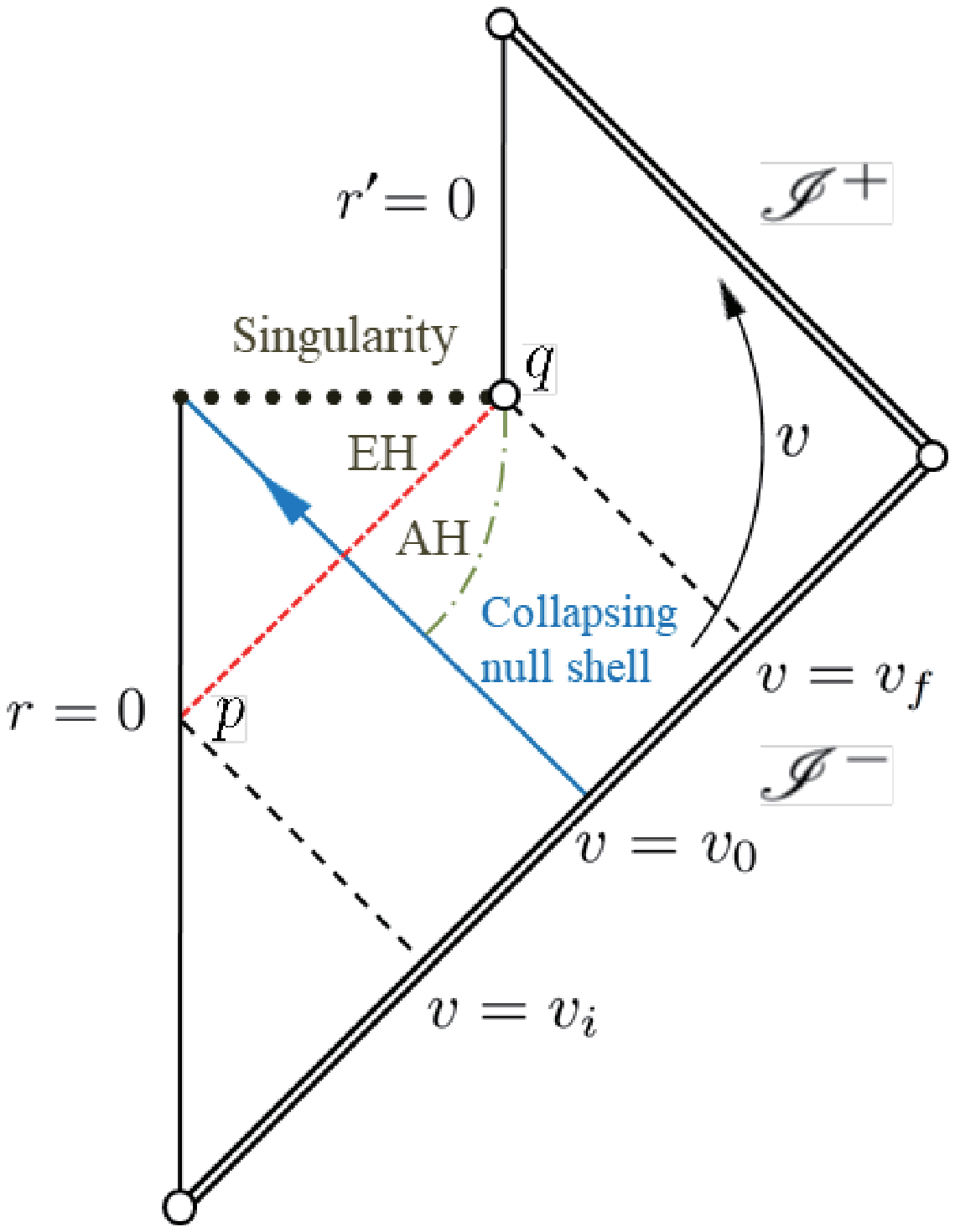}
 \end{center}
 \caption{The Penrose diagram of a four-dimensional spherically symmetric evaporating black hole which is formed by a collapsing null 
 dust shell. 
The (red) dotted line and the (green) dashed-dotted line represent, respectively, the event horizon and the apparent horizon. The (blue) 
solid line represents the collapsing null dust shell, which is characterized in terms of the advanced null coordinate $v$ as $v=v_0$. }
 \label{fig:two}
\end{figure}

In this section, to examine the ANEC, we construct the 5-dimensional bulk geometry dual to a $4$-dimensional asymptotically 
flat, spherically symmetric time-dependent boundary black hole, describing the formation by gravitational collapse and subsequent evaporation. 
Such a bulk spacetime can be constructed, at least locally near the boundary, 
by taking the Vaidya metric as our boundary geometry and performing the Fefferman-Graham expansion into $5$-dimensional bulk. 
As such, although our construction of the holographic time-dependent black hole is limited, it is sufficient for the purpose of 
examining the ANEC on our time-dependent boundary black hole by applying the 
method of Refs.~\cite{IizukaIshibashiMaeda2020, IizukaIshibashiMaeda2020_Ver2}, based on the no-bulk shortcut principle.

\subsection{An asymptotically flat evaporating black hole}

We consider an asymptotically flat spherically symmetric evaporating black hole formed by injecting a null dust shell. Such a model of evaporating black hole is well described by the Vaidya model~\cite{Hiscock1981} as,  
\begin{align} 
\label{evaporateBH_metric}
ds^2=g_{\mu \nu}dx^\mu dx^\nu =2dvdr-\left(1-\frac{2m(v)}{r}  \right)dv^2+r^2(d\theta^2+\sin^2\theta d\phi^2) \,, 
\end{align}
where $v$ denotes the advanced null coordinate and $m(v)$ the mass of the black hole. 
Note that here and hereafter, we use the coordinate index $x^\mu$ to denote tensor components on the $4$-dimensional boundary spacetime. 
As shown in Fig.~\ref{fig:two}, the null dust shell is injected at $v=v_0$ to form the black hole, of which the event horizon~(EH) begins 
at the center of spherical symmetry~[point $p$~at ($r=0, v=v_i$)] and terminates at a zero mass naked singularity at point $q$~($r=0, v=v_f$), where 
the evaporation is completed. The apparent horizon~(AH) is located outside the EH during the evaporating process, 
as the area decreases. We assume that the spacetime is flat, i.e., $m(v)=0, v<v_0$ before the collapse of the null shell.

As is often discussed in the consideration of backreaction effects of the Hawking radiation, for a free massless field, the negative energy flux ${\cal F}$ flowing into the black hole is determined by the dimensional analysis as 
\be 
{\cal F}\simeq \frac{A}{m(v)^2} \,,  
\ee
where $A$ is of order unity in Planck units~\cite{Wald_book1994}. Assuming that the mass of the 
black hole is lost by the negative energy flux, ${dm}/{dv} \simeq -A/m^2$, one obtains 
\be 
\label{mass_loss_ratio}
m(v)\sim (v-v_f)^{1/3}  \,. 
\ee
Generalizing Eq.~(\ref{mass_loss_ratio}), we shall consider a holographic model in which 
the AdS boundary is conformal to the Vaidya metric~(\ref{evaporateBH_metric}) with $m(v)$ satisfying 
\begin{align}
\label{mass_function}
m(v)=m_0\left(1-\frac{v_0}{v_f}\right)^{-\alpha}\left(1-\frac{v}{v_f}\right)^\alpha \,, \qquad m_0>0 \,, \quad 0<\alpha<1 \,,  
\end{align} 
where $m_0$ is the mass of the black hole just after the collapse of the null dust shell. 
The radial null geodesic on the EH obeys the equation  
\begin{align}
\label{null_geodesic}
\frac{dr}{dv}=\frac{1}{2}\left(1-\frac{2m(v)}{r} \right)\,. 
\end{align}
Note that the AH is located at the radius $r_{\rm AH}$ where the r.h.s. of the above equation vanishes, i.e., $r_{\rm AH}=2m(v)=2m_0(1-v/v_f)^\alpha$. 
In the evaporation process $v_0<v<v_f$, the EH is located inside the AH as depicted in Fig.~\ref{fig:two}, which implies 
that the EH radius $r_{\rm EH}$ is bounded from above by $r_{\rm AH}=2m(v)$. 
This can be the case since the quantum field stress-energy will not satisfy the weak energy condition under the evaporation process. 
In particular, at the last stage of the evaporation, the geometry rapidly changes and it is plausible that the EH radius could 
be much smaller than that of the AH, close to the final point of the evaporation $v\simeq v_f$. 
In such a case, by assuming $r_{\rm EH}\ll 2m(v)$, we can find the approximated solution of (\ref{null_geodesic}) near $v=v_f$ as,    
\begin{align}
\label{singular_r}
r_{\rm EH}(v) \simeq \sqrt{\frac{2m_0 v_f}{\alpha+1}}\left(1-\frac{v_0}{v_f}\right)^{-\frac{\alpha}{2}}
\left(1-\frac{v}{v_f}\right)^{(\alpha+1)/2} \,. 
\end{align}

Along the radial outgoing null vector  
\begin{align}
\label{tangentNull}
k^\mu= (\p_v)^\mu + \dfrac{1}{2}\left(1-\dfrac{2m(v)}{r}  \right) (\p_r)^\mu \,, 
\end{align}
on the EH, the affine parametrized null geodesic generator $\gamma$ is given by 
\be 
\label{affine_Null}
K^\mu:=h(v) k^\mu \,,  
\ee
where the factor $h$ obeys the following equation,  
\begin{align}
\label{Eq_h}
\frac{dh}{dv}=- {h}\frac{m}{r^2} \,. 
\end{align}
Near the naked singularity, $v=v_f$, $r=0$, the solution is approximately given by the functions~(\ref{mass_function}) and 
(\ref{singular_r}) as
\begin{align}
\label{singular-h}
\frac{dv}{dV}=h \propto \left(1-\frac{v}{v_f}\right)^{(\alpha+1)/2} \,,   
\end{align}
where $V$ is the affine parameter of the null geodesic $\gamma$ with the tangent $K^\mu$.

\subsection{The no bulk-short cut principle in the dual bulk spacetime}
We briefly recall the statement of the no-bulk-short-cut principle proposed 
in~\cite{KellyWall2014, IizukaIshibashiMaeda2020, IizukaIshibashiMaeda2020_Ver2}. 
Suppose that a pair of boundary points $p$ and $q$ are connected by an {\em achronal} null geodesic $\gamma$ 
on the boundary spacetime. Then, the no-bulk-short-cut principle states that 
there is no bulk timelike curve $\lambda$ which connects $p$ to $q$. In other words, $\gamma$ is the {\it fastest} causal 
curve among all causal curves from $p$ to $q$, including the bulk causal curves. In a class of spacetime, this principle 
was shown in the Gao-Wald theorem~\cite{GaoWald2000}. By applying this principle to boundary Minkowski spacetime in a holographic setting, 
the ANEC was proved in \cite{KellyWall2014}. The principle is also applied to the spatially compact 
universes and the CANEC or the ANEC with a weight function were derived~\cite{IizukaIshibashiMaeda2020, IizukaIshibashiMaeda2020_Ver2}. In this regard, 
see also \cite{Rosso} for the ANEC in de Sitter and AdS spacetime.
Here, we shall apply this principle to the holographic model with the boundary geometry descrbing the evaporating black hole  (\ref{evaporateBH_metric}) discussed above, whose event horizon is generated by achronal null geodesics $\gamma$ from $v=v_i$ to $v=v_f$. 
We wish to derive the lower bound of the null-null component of the boundary stress-energy tensor along boundary achronal
null geodesics $\gamma$. Our strategy is very similar to the one performed
in Refs.~\cite{IizukaIshibashiMaeda2020, IizukaIshibashiMaeda2020_Ver2}, which is as follows. Given such a boundary achronal 
null geodesic $\gamma$, we consider all the bulk causal curves connected with $\gamma$ via Jacobi fields,
thus considered to be in the neighborhood of $\gamma$ covered by Fefferman-Graham coordinates.
Then, by imposing the no-bulk-short-cut principle that $\gamma$ be the fastest causal curve
among all the causal curves in the neighborhood, we will obtain the lower bound. 
 
The $5$-dimensional bulk spacetime in the neighbourhood of the boundary metric~(\ref{evaporateBH_metric}) including the boundary 
null geodesics is represented by the following Fefferman-Graham coordinates 
\begin{align}
\label{FG_coordinate_five}
& g_{ab} dx^a dx^b=\frac{L^2}{z^2} \left( dz^2+g_{\mu\nu}(x,z)dx^\mu dx^\nu \right) 
:= \frac{L^2}{z^2} \hat{g}_{ab}dx^a dx^b \,, \nonumber \\
& g_{\mu\nu}(z,x)=g_{(0)\mu\nu}(x)+z^2g_{(2)\mu\nu}(x)+\cdots + z^4 g_{(4)\mu\nu}(x)+h_{(4)\mu\nu}\,z^{4}\ln z^2+\cdots \,, 
\end{align}
where $g_{(0)\mu\nu}(x)$ corresponds to the boundary metric~(\ref{evaporateBH_metric}), and 
$\hat{g}_{ab}$ is the unphysical metric conformal to $g_{ab}$. The subleading coefficients 
$g_{(2)\mu\nu}$, $g_{(4)\mu\nu}(x)$, and $h_{(4)\mu\nu}$ are given by~\cite{HaroSkenderisSolodukhin2001}, 
\begin{align}
\label{coefficient_FG_five}
& g_{(2)\mu\nu}=-\frac{1}{2}\left(R_{\mu\nu}-\frac{R}{6}g_{(0)\mu\nu}\right) \,, \nonumber \\
& g_{(4)\mu\nu}=t_{\mu\nu}+\frac{1}{8}g_{(0)\mu\nu}\left[(\mbox{Tr} g_{(2)})^2-\mbox{Tr}(g^2_{(2)})   \right]
+\frac{1}{2}g_{(2)\mu\alpha}{g_{(0)}}^{\alpha\beta}g_{(2)\beta\nu}-\frac{1}{4}g_{(2)\mu\nu}\mbox{Tr}(g_{(2)}) \,, 
\nonumber \\
& h_{(4)\mu\nu}=\frac{1}{2}g_{(2)\mu\alpha}{g_{(0)}}^{\alpha\beta}g_{(2)\beta\nu}
-\frac{1}{8}g_{(0)\mu\nu}\mbox{Tr}(g_{(2)}^2) \,,  
\end{align}
where $R_{\mu\nu}$, $R$ are the Ricci tensor, Ricci scalar of the metric $g_{(0)\mu\nu}$, and the indices are raised and 
lowered by the metric $g_{(0)\mu\nu}$. According to the AdS/CFT dictionary~\cite{HaroSkenderisSolodukhin2001}, 
the expectation value of the stress-energy tensor $\Exp{T_{\mu\nu}}$ is determined by 
$t_{\mu\nu}$ as 
\begin{align}
\label{def:t_munu_d=4}
\Exp{T_{\mu\nu}}=\frac{1}{4\pi G_5}t_{\mu\nu} \,.
\end{align} 

It is convenient to use the following double null coordinates  
\begin{align}
\label{boundary:d=4} 
g_{(0)\mu\nu}dx^\mu dx^\nu = -e^{f(U,V)}dUdV+r^2(U,V)(d\theta^2+\sin^2\theta d\varphi^2) \,, 
\end{align}  
instead of the metric~(\ref{evaporateBH_metric}). On the EH, we can set 
\be
U=0 \,, \quad f(0,V)=0 \,,   
\ee
wthout loss of generality. In this case, $V$ is the affine parameter of the boundary null geodesic $\gamma$ with the tangent vector $K^\mu$ defined by Eq.~(\ref{affine_Null}). We denote $V_i$ and $V_f$ as the 
values of the point $p$~($v=v_i$) and the point $q$~($v=v_f$), respectively.  

Now consider a bulk causal curve $\lambda$ with the tangent vector $K^a$ near the boundary 
null geodesic $\gamma$, which connects the boundary point $p$ to the boundary point $q$ on $\gamma$. 
From spherical symmetry, it is sufficient to consider the case $K^\theta=K^\phi=0$, and $K^a$ can be expanded as a series of small 
parameter $\epsilon$ as 
\begin{align} 
& K^a=(K^z, K^U, K^V, K^\theta, K^\phi)=
\left(\frac{dz}{dV}, \frac{dU}{dV}, 1, 0,0  \right) \,, 
\nonumber \\
& z=\epsilon z_1+\epsilon^2z_2+\cdots \,, 
\nonumber \\
& \frac{dU}{dV}=\epsilon^2\frac{dU_2}{dV}+\epsilon^3\frac{dU_3}{dV}+\epsilon^4\frac{dU_4}{dV}+\cdots \,. 
\end{align}
Substituting this into Eq.~(\ref{FG_coordinate_five}), the constraint of $\hat{g}_{ab} K^a K^b\le 0$ is expressed as  
\begin{align}
\label{Causal_exp_2}
\hat{g}_{ab}K^a K^b 
& = \epsilon^2\left(-\frac{dU_2}{dV}+\dot{z}_1^2+z_1^2\,g_{(2)VV}(0,V)\right)   \nonumber \\
&\quad +\epsilon^3\left(2\dot{z}_1\dot{z}_2+2z_1z_2\,g_{(2)VV}(0,V)-\frac{dU_3}{dV}  \right) \nonumber \\
&\quad +\epsilon^4\Biggl(\dot{z}_2^2+z_2^2\,g_{(2)VV}(0,V)+2\dot{z}_1\dot{z}_3 \nonumber \\
&\qquad \qquad +2z_1z_3\,g_{(2)VV}(0,V) +2z_1^4\ln z_1\,h_{VV}(0,V) \nonumber \\
&\qquad \qquad +2z_1^2\,g_{(2)UV}(0,V)\frac{dU_2}{dV}-\frac{dU_4}{dV}+z_1^4\,g_{(4)VV}(0,V) \nonumber \\
&\qquad \qquad +z_1^2\p_U(g_{(2)VV})(0,V)U_2-(\p_U f)(0,V)U_2\frac{dU_2}{dV} \Biggr) \nonumber \\
&\quad + 2\epsilon^4\ln (\epsilon) z_1^4\,h_{VV}(0,V)+\cdots \nonumber \\ 
& \quad \le 0 \,,  
\end{align}
where the dot denotes the derivative with respect to $V$. 
At $O(\epsilon^2)$, we obtain the inequality of the variation of $U$ from $V_i$ to $V_f$ as  
\begin{align}
\label{second_variation_U}
\Delta U_2 = \int_{V_i}^{V_f} \dfrac{dU_2}{dV} dV  \ge \int^{V_f}_{V_i}(\dot{z}_1^2+z_1^2 \,g_{(2)VV}(0,V))dV \, ,  
\end{align}
where the equality is satisfied for the null curve. The r.h.s. of Eq.~(\ref{second_variation_U}) is minimized for $z_1$ obeying 
\begin{align}
\label{z_1_jacobi}
\ddot{z}_1=g_{(2)VV}(0,V)z_1=\frac{\ddot{r}(0,V)}{r(0,V)}z_1 \,,  
\end{align} 
where we have used that on the EH, $R_{VV}(0,V)= -2\ddot{r}/r$, under our gauge condition $f(0,V)=0$ and the holographic formula (\ref{coefficient_FG_five}) for $g_{(2)VV}$. 
Therefore the solution satisfying $z_1(p)=z_1(q)=0$ is simply given by 
\begin{align}
\label{sol:z1}
z_1= c r(0,V) \,, 
\end{align}
for some positive constant $c$. Then, the integrand of the r.h.s. of (\ref{second_variation_U}) is expressed simply as $(z_1 \dot{z}_1)\dot{}$ and thus, 
integrating the r.h.s. of Eq.~(\ref{second_variation_U}) results in the boundary values of $z_1 \dot{z}_1$. 
As shown in Ref.~\cite{IizukaIshibashiMaeda2020_Ver2}, $z_1$ is identified as the magnitude of a Jacobi 
field for a null geodesic congruence on EH. Then, $z_1$ must vanish at $(U=0, V=V_i)$ [the point $p$ in Fig.~\ref{fig:two})], 
where the EH is formed. Also, since the EH completely evaporates at $(U=0,V=V_f)$ [the point $q$ in Fig.~\ref{fig:two})], the Jacobi field $z_1$ is expected to vanish there too; 
\begin{align}
z_1(p)=z_1(q)=0 \,.  
\end{align}
This guarantees that the bulk null curve $\lambda$ remains in the neighbourhood of the boundary null geodesic $\gamma$ from 
the point $p$ to the point $q$ when $\epsilon$ is small enough. 
In fact, one can directly check that the boundary value $\dot{z}_1z_1(V)$ goes to zero in the limit $V\to V_f$ by utilizing Eqs.~(\ref{singular_r}) and 
(\ref{singular-h}), as  
\begin{align}
\lim_{V\to V_f}\dot{z}_1(V)z_1(V)=\lim_{V\to V_f}h(v)\frac{dz_1}{dv}z_1
\sim \lim_{v\to v_f}\left(1-\frac{v}{v_f}\right)^{(3\alpha+1)/2}\to 0 \,.  
\end{align} 
Thus, the r.h.s. of Eq.~(\ref{second_variation_U}) must vanish, implying that 
there is no time delay between $\gamma$ and $\lambda$ at $O(\epsilon^2)$ when $\lambda$ is the null curve, as the variation of $U$ 
between $\gamma$ and $\lambda$ is zero. 
Actually, $U_2$ is analytically written by $z_1$ as 
\begin{align}
\label{sol:U_2}
U_2=z_1\dot{z}_1 \,. 
\end{align} 

Similarly, the variation of $U$ from $V_i$ to $V_f$ is minimized for $z_2$ and $z_3$ obeying the same Jacobi equation~(\ref{z_1_jacobi}), and 
$\Delta U_3$ is also zero at $O(\epsilon^3)$. 

As for $O(\epsilon^4)$, by noting the fact that for the Vaidya metric~(\ref{evaporateBH_metric}), only non-vanishing 
component among $g_{(2)\mu \nu}$ in Eqs.~(\ref{coefficient_FG_five}) is $g_{(2) VV} = -R_{VV}/2$. Thus, $h_{VV}=0$, and we obtain 
\begin{eqnarray}
\label{four_U_4}
\Delta U_4 &\ge& \int^{V_f}_{V_s}z_1^4\,t_{VV}(0,V)dV+{\cal I} \,,  
\nonumber \\ 
{\cal I} &:=&\int^{V_f}_{V_s}\left(z_1^2\p_U(g_{(2)VV})(0,V)U_2-(\p_U f)(0,V)U_2\frac{dU_2}{dV}\right)dV \nonumber \\
          &=& \frac{c^4}{16} \int^{V_f}_{V_i} r^2\left[r^2\mu \,\theta_+^2-2R_{VV} \right] dV \,, 
\end{eqnarray} 
where $\theta_\pm$ and $\mu$ are outgoing and ingoing expansions, and the mass density defined by 
\begin{align}
\label{def:theta}
\theta_+=\frac{2\dot{r}}{r} \,, \quad 
\theta_-=\frac{2r'}{r} \,, 
\quad 
\mu:=\frac{1}{r^2}+\theta_+\theta_- \,, 
\end{align}
where the dash expresses the derivative with respect to $U$. 
The second equality in Eq.~(\ref{four_U_4}) is shown in the Appendix. 
Applying the no-bulk-short-cut principle to our case,  $\Delta U_4$ should be non-negative, and we finally obtain the 
following inequality with the weight function $r^4$, 
\begin{align}
\label{inequality_EvapolatingBH}
\int^{V_f}_{V_s} r^4 t_{VV}dV\ge \frac{1}{16}\int^{V_f}_{V_s} r^2\left[2R_{VV}-r^2\mu \,\theta_+^2   \right]dV \,. 
\end{align}

\subsection{The ANEC in the evaporating BH}
Now, we can derive the ANEC from Eq.~(\ref{inequality_EvapolatingBH}). Using the fact that $f(0,V)=0$ and $\p_U=-1/(2h)\p_r$ 
along the EH, the expansions $\theta_\pm$ and the mass density $\mu$ in Eqs.~(\ref{def:theta}) are rewritten 
in terms of the advanced null coordinate, $v$ and $r$ as 
\begin{align}
\label{v_r_expansion}
\theta_+=\left(1-\frac{2m(v)}{r}  \right)\frac{h}{r}, \qquad \theta_-=-\frac{1}{hr}, \qquad \mu=\frac{2m}{r^3}. 
\end{align}
Substituting Eq.~(\ref{v_r_expansion}) into the inequality~(\ref{inequality_EvapolatingBH}), we find 
\begin{align}
\label{inequality_EvapolatingBH1}
\int^{V_f}_{V_s} r^4 t_{VV}dV\ge \frac{1}{4}\int^{v_f}_{v_s}h\left[\frac{dm}{dv}-\left(1-\frac{2m}{r(v)}\right)^2\frac{m}{2r(v)}  \right]dv. 
\end{align}

In order to inspect the above inequality closely, let us devide the whole region $v_i\le v\le v_f$ into four phases: 

\medskip 

{Phase (I)} $v_i\le v\le v_0-\delta_1$: the geometry is the Minkowski spacetime, 

{Phase (II)} $v_0-\delta_1\le v\le v_0+\delta_1$: the null shell is collapsing, 

{Phase (III)} $v_0+\delta_1\le v\le v_f-\delta_2$: the BH is evaporating,  

{Phase (IV)} $v_f-\delta_2\le v\le v_f$: the final explosion. 

\medskip  
\noindent 
Here, we assume that the time evolution of the evaporation is sufficiently slow~($v_f\gg m_0$), and the time durations $\delta_1$, $\delta_2$ 
are very small, i.e., 
\be
\label{def:delta_12}
\delta_1, \,\delta_2 \ll |v_f-v_0| \,.  
\ee
In Phase (I), the r.h.s. of Eq.~(\ref{inequality_EvapolatingBH1}) vanishes, as $m=0$. In the collapsing phase (II), 
the mass rapidly grows during the short time period $\delta_1$ by the condition~(\ref{def:delta_12}), and the second term is negligible compared with the first term. In Phase (III), the time evolution of the mass is very slow, and $r\simeq 2m$ in this phase. 
Let us set the location of the EH as $r=2m (1-\eta) $ with $\eta$ assumed to be small so that $|\eta| \simeq |dm/dv |\ll 1$ in Phase (III). 
Then, linearizing Eq.~(\ref{null_geodesic}) with respect to $\eta$, one obtains the approximate solution,   
\begin{align}
\label{app_sol_xi}
 \eta \simeq -2 \dfrac{d}{dv}[2m(1-\eta)] \simeq - 4\dfrac{dm}{dv} \,.
\end{align}
This implies that the second term in the r.h.s. of Eq.~(\ref{inequality_EvapolatingBH1}) is small compared with the first term, as  
\begin{align}
\left(1-\frac{2m}{r}\right)^2\frac{m}{2r}\simeq \frac{\eta^2}{4}\ll \Bigl{|}\frac{dm}{dv}\Bigr{|}\simeq \Bigl{|}\frac{\eta}{4}\Bigr{|} \,, 
\end{align}
under the approximation~(\ref{app_sol_xi}). 

In the final phase (IV), one might expect that the second term in the r.h.s. of Eq.~(\ref{inequality_EvapolatingBH1}) could give a large contribution. However, in the case that the time evolution of the mass is very slow, i.e., $v_f\gg m_0$, it turns out that the contribution from the second term is still small compared with that from the first term in Phase (IV). This can be seen from Eq.~(\ref{singular_r}) that the ratio of the second term to the first term is at least of $O( \sqrt{m_0/v_f})$. 
%
%
Thus, it is sufficient for our purpose to evaluate the first term of the r.h.s. of Eq.~(\ref{inequality_EvapolatingBH1}) in each phase. 
In Phase (I), the spacetime is flat, and then $h=1$. In Phase (II), by integrating Eq.~(\ref{Eq_h}), we obtain 
\begin{align}
\int^{v_0+\delta_1}_{v_0-\delta_1} \frac{dh}{h}=\ln h(v_0+\delta_1) =-\int^{v_0+\delta_1}_{v_0-\delta_1}\frac{m}{r^2}dv=O(\delta_1) \,,  
\end{align}
which implies $h\sim 1$ during Phase (II). 
Thus, we obtain 
\begin{eqnarray}
\label{evaluate_In}
 \int^{v_f}_{v_i}h\frac{dm}{dv}dv
 &=& \int^{v_0+\delta_1}_{v_0-\delta_1}h\frac{dm}{dv}dv+\int^{v_f}_{v_0+\delta_1}h\frac{dm}{dv}dv
\nonumber \\
  &\simeq& \int^{v_0+\delta_1}_{v_0-\delta_1}\frac{dm}{dv}dv+\int^{v_f}_{v_0+\delta_1}h\frac{dm}{dv}dv \nonumber \\
\nonumber \\  
 &=& m_0+\int^{v_f}_{v_0+\delta_1}h\frac{dm}{dv}dv \,.  
\end{eqnarray}
By Eq.~(\ref{Eq_h}), $dh/dv<0$ during the evaporating phases (III) and (IV), 
\be 
0\le h\le 1 \quad \mbox{for} \quad v\ge v_0+\delta_1 \,.  
\ee
Since $dm/dv<0$  during the evaporating phases (III) and (IV), we thus obtain 
\begin{eqnarray}
 \int^{V_f}_{V_s} r^4 t_{VV}dV &\ge& \frac{1}{4}\left[m_0+\int^{v_f}_{v_0+\delta_1}h\frac{dm}{dv}dv \right] 
 \nonumber \\ 
 &\ge&  
 \frac{1}{4}\left[m_0+\int^{v_f}_{v_0+\delta_1}\frac{dm}{dv}dv \right] 
 \nonumber \\
 &\ge& \frac{1}{4}m_0+\frac{1}{4}[m(v)]^{v_f}_{v_0+\delta_1}
 \nonumber \\
 &\ge &0 \,.
 \end{eqnarray} 
This shows that the ANEC with the weight function $r^4$ is satisfied for the evaporating black hole.

\section{Summary and discussions}
\label{sec:4}
We have investigated two holographic models of evaporating black holes. First, by perturbing the 4-dimensional black droplet 
solution associated with BTZ black holes on the AdS boundary~\cite{HubenyMarolfRangamani2010_Ver2}, we have constructed the bulk 
geometry dual to the boundary field theory in a time-dependent BTZ type black hole with the horizon area decreasing. 
We found that negative energy flux going into the time-dependent horizon always appears by calculating the boundary stress-energy 
tensor. This calculation agrees with Hawking's picture of evaporating black holes~\cite{Hawking1975}, where the 
horizon area decreases by absorbing negative quantum energy flux. According to the energy conservation, 
the total energy and the matter entropy outside the evaporating black hole should increase by the Hawking radiation. 
Our result supports the generalized second law~(GSL)~\cite{Hawking1975,Bekenstein1973}, which states that 
the total of the gravitational entropy and the matter entropy outside the horizon should not decrease. 

The existence of the negative energy flux shown in this paper implies a local violation of the null energy condition~(NEC). 
As first shown in a class of holographic theories~\cite{KellyWall2014}, 
ANEC is satisfied along a complete achronal null geodesic in flat spacetimes. 
In curved spacetime, however, we cannot expect that the ANEC is generically satisfied, 
except some particular class of curved spacetimes~\cite{IizukaIshibashiMaeda2020, IizukaIshibashiMaeda2020_Ver2}. 
It was shown for some semi-analytic black droplet solutions~\cite{Haddad2012, IMM2017} that 
 along the incomplete null geodesics, which terminate at a singularity inside the boundary black hole, the ANEC is violated due to the negative divergence of the null energy toward the singularity. 
In our present holographic model, 
the null geodesic generators along the event horizon begin from the formation of the boundary black hole, and terminate 
at the final instant of the black hole evaporation, where a naked singularity appears, as shown in Fig.~\ref{fig:two}. Although 
this situation is very similar to the incomplete null geodesics observed in Refs.~\cite{Haddad2012, IMM2017}, we can 
expect the ANEC to hold in our case, as the singularity at the very final moment of the complete black hole evaporation is a {\it weak} singularity in the sense 
that the mass $m(v)$ becomes zero at the final singularity. In addition, the null geodesic congruence along the horizon expands from 
the point at the formation of the black hole, and shrinks to zero again at the final zero mass singularity. 
This behavior of the null geodesic congruence is very similar to that in a spatially compact $S^3$ universe, in which a 
congruence expands from the south pole and then shrinks to zero at the north pole. This peculiar feature enables us 
to prove the ANEC along the null geodesic generator of the event horizon, just like the case considered in spatially  
compact universes~\cite{IizukaIshibashiMaeda2020, IizukaIshibashiMaeda2020_Ver2}. This indicates that the total negative 
energy flux is bounded from below, and is compensated by a positive energy flux in the early stage of the black hole 
formation.

\section*{Acknowledgements}
We would like to thank Takashi~Okamura for useful discussions. This work was supported in part by 
JSPS KAKENHI Grant Numbers 17K05451, 20K03975~(KM), and 15K05092, 20K03938, 21H05182~(AI), 
and 21H05186 (KM, AI).

\appendix
\section{The derivation of Eq.~(\ref{inequality_EvapolatingBH})}
In the Vaidya metric~(\ref{evaporateBH_metric}), $R_{\mu\nu}=0$ except $R_{vv}$. In the double null coordinates~(\ref{boundary:d=4}), 
$R_{UV}=R_{\theta\theta}=0$ yields  
\begin{align}
\label{Eq:k_r}
\dot{f'}=-\dfrac{2\dot{r'}}{r}, \qquad \dot{r'}=-\dfrac{1}{r}\left(\dot{r}r'+\frac{1}{4}e^{f}\right). 
\end{align}
By the fact that $f=0$ and $\p_U\sim h^{-1}\p_r$ along the EH where $U=0$, the asymptotic behavior of $\dot{r'}$ and $f'$ in 
Eq.~(\ref{Eq:k_r}) becomes 
\be 
\label{singular_k'_dot_r'}
\dot{r'}\sim \left(1-\dfrac{v_f}{v}\right)^{-1}, \quad f'\sim \left(1-\dfrac{v_f}{v}\right)^{-\alpha-1}
\ee
from Eqs.~(\ref{mass_function}), (\ref{singular_r}), and (\ref{singular-h}). 

By Eq.~(\ref{coefficient_FG_five}), $g_{(2)VV}$ is given by 
\be 
\label{Ricci_VV_double}
g_{(2)VV}=-\dfrac{R_{VV}}{2}=\dfrac{\ddot{r}-\dot{f}\dot{r}}{r}. 
\ee
Substituting this into the first term of ${\cal I}$ in Eq.~(\ref{four_U_4}), we obtain 
\be
\label{action_I_1}
 \int z_1^2\p_U(g_{(2)VV})U_2\,dV
&=&\int z_1^3\dot{z}_1
\left[\dfrac{\ddot{r'}}{r}-\frac{r'}{r^2}\ddot{r}-\dfrac{2\dot{r}}{r^3}\left(\dot{r}r'+\dfrac{1}{4}\right)   \right]dV
\nonumber \\
&=&c^4\int \dot{r}
\left[r^2\ddot{r}'-rr'\ddot{r}-2\dot{r}\left(\dot{r}r'+\dfrac{1}{4}\right) \right]dV \nonumber \\
&=&\frac{c^4}{4}\int r\ddot{r}dV.  
\ee
Here, in the second line, we used the condition $f(0,V)=0$ along the EH and also Eqs.~(\ref{sol:U_2}), (\ref{Eq:k_r}), and (\ref{sol:z1}).  
In the third line, we performed integration by parts for the 
first term in the second line, and used the fact that the boundary term $r^2\dot{r'}\dot{r}$ vanishes in the 
limit $v\to v_f$ by Eqs.~(\ref{singular_r}), (\ref{singular-h}), and (\ref{singular_k'_dot_r'}).  

Similarly, we evaluate the second term of ${\cal I}$ in Eq.~(\ref{four_U_4}) as 
\begin{align}
\label{action_I_2}
-\int f'U_2\dot{U}_2dV
=-\int \dfrac{\dot{r'}}{r}U_2^2dV
= c^4\int \left(\dot{r}r'+\dfrac{1}{4}  \right)\dot{r}^2dV, 
\end{align}
where we performed integration by parts in the first equality and used Eq.~(\ref{Eq:k_r}). 
Note that the boundary term $f'U_2^2$ vanishes in the limit $v\to v_f$ by 
Eqs.~(\ref{singular_r}), (\ref{singular-h}), and (\ref{singular_k'_dot_r'}).   
To derive the second equality, we used Eqs.~(\ref{singular_k'_dot_r'}) and (\ref{sol:z1}). 
Substituting Eqs.~(\ref{action_I_1}) and (\ref{action_I_2}) into ${\cal I}$ in Eq.~(\ref{four_U_4}) and using 
Eqs.~(\ref{v_r_expansion}) and (\ref{Ricci_VV_double}), we finally obtain the inequality~(\ref{inequality_EvapolatingBH}).



\begin{thebibliography}{99}
\bibitem{Hawking1975}
S.~Hawking, Particle Creation by Black Holes, Commun.~Math.~Phys. {\bf 43} (1975) 199. 
\bibitem{DaviesFullingUnruh1976}
P.~C.~W.~Davies, S.~A.~Fulling, and W.~G.~Unruh, Energy-momentum tensor near an evaporating black hole, 
Phys.~Rev.~{\bf D13} (1976) 2720. 
\bibitem{Unruh1976}
W.~G.~Unruh, Notes on black hole evaporation, Phys.~Rev.~{\bf D14} (1976) 870. 
\bibitem{Maldacena1998}
J.~M.~Maldacena, The Large N limit of superconformal field theories and supergravity,
Adv.~Theor.~Math.~Phys. {\bf 2} (1998) 231, arXiv:hep-th/9711200.
\bibitem{FiguerasLuciettiWiseman2011}
P.~Figueras, J.~Lucietti, and T.~Wiseman,  
Ricci solitons, Ricci flow, and strongly coupled CFT in the Schwarzschild Unruh or Boulware vacua, 
Class.~Quant.~Grav. {\bf 28} (2011)  215018, arXiv:1104.4489 [hep-th].
\bibitem{SantosWay2012}
J.~E.~Santos and B.~Way, Black Funnels, JHEP {\bf 12} (2012) 060, arXiv:1208.6291 [hep-th].
\bibitem{FischettiSantos2013}
S.~Fischetti and J.~E. ~Santos, Rotating Black Droplet, 
JHEP {\bf 07} (2013) 156 arXiv:1304.1156 [hep-th]. 
\bibitem{FT2013}
P.~Figueras and S.~Tunyasuvunakool, 
CFTs in rotating black hole backgrounds, 
Class.~Quant.~Grav. {\bf 30} (2013) 125015 arXiv:1304.1162 [hep-th].
\bibitem{Mefford2017}
E.~Mefford, Entanglement Entropy in Jammed CFTs, 
JHEP {\bf 09} (2017) 006 arXiv:1605.09369 [hep-th]. 
\bibitem{HubenyMarolfRangamani2010}
V.~E.~Hubeny, D.~Marolf, and M.~Rangamani, 
Hawking radiation in large N strongly-coupled field theories,  
Class.~Quant.~Grav. {\bf 27} (2010) 095015, arXiv:0908.2270 [hep-th]. 
\bibitem{FischettiMarolf2012}
S.~Fischetti and D.~Marolf, Flowing Funnels: Heat sources for field theories and the AdS$_3$ dual of CFT$_2$ Hawking radiation, 
Class. Quant. Grav. {\bf 29} (2012) 105004, arXiv:1202.5069 [hep-th]. 
\bibitem{FischettiMarolfSantos2013}
S.~Fischetti, D.~Marolf, and J.~E. Santos, 
AdS flowing black funnels: Stationary AdS black holes with non-Killing horizons and heat transport in the dual CFT, 
Class.~Quant.~Grav. {\bf 30} (2013) 075001, arXiv:1212.4820 [hep-th]. 
\bibitem{Bekenstein1973}
J.~D.~Bekenstein, Black holes and entropy, Phys.~Rev.~{\bf D7} (1973) 2333. 
\bibitem{Wall2012}
A.~C.~Wall, A proof of the generalized second law for rapidly changing fields and arbitrary horizon slices, 
Phys.~Rev.~{\bf D85} (2012) 104049, arXiv:1105.3445 [hep-th].  
\bibitem{BuntingFuMarolf2016}
W.~Bunting, Z.~Fu, and D.~Marolf, A coarse-grained generalized second law for holographic conformal field theories, 
Class. Quant. Grav. {\bf 33} (2016) 055008, arXiv:1509.00074 [hep-th]. 
\bibitem{HubenyMarolfRangamani2010_Ver2}
V.~E. Hubeny, D.~Marolf, and M.~Rangamani, 
Hawking radiation from AdS black holes, 
Class.~Quant.~Grav. {\bf 27} (2010) 095018, arXiv:0911.4144 [hep-th]. 
\bibitem{BTZblackhole}
M.~Banados, C.~Teitelboim, and J.~Zanelli, Black hole in three-dimensional spacetime, 
Phys. Rev. Lett. {\bf 69} (1992) 1849, arXiv:hep-th/9204099.  
\bibitem{IizukaIshibashiMaeda2020}
N.~Iizuka, A.~Ishibashi, and K.~Maeda, Conformally invariant averaged null energy condition from AdS/CFT, 
JHEP {\bf 03} (2020) 161 arXiv:1911.02654 [hep-th].
\bibitem{IizukaIshibashiMaeda2020_Ver2}
N.~Iizuka, A.~Ishibashi, and K.~Maeda, The averaged null energy conditions in even dimensional curved spacetimes 
from AdS/CFT duality, JHEP {\bf 10} (2020) 106 arXiv:2008.07942 [hep-th].
\bibitem{GaoWald2000}
S.~Gao and R.~M.~Wald, Theorems on gravitational time delay and related issues, 
Class.~Quant.~Grav. {\bf 17} (200) 4999, arXiv:gr-qc/0007021.
\bibitem{HaroSkenderisSolodukhin2001}
S.~d.~Haro, K.~Skenderis, and S.~N.~Solodukhin, Holographic reconstruction of spacetime and renormalization in the
AdS/CFT correspondence, Comm.~Math.~Phys. {\bf 217} (2001) 595,  arXiv:hep-th/0002230.  
\bibitem{KodamaIshibashi2003}
H.~Kodama and A.~Ishibashi, A master equation for gravitational perturbations of maximally symmetric black holes in higher dimensions, 
Prog.~Theor.~Phys. {\bf 110} (2003) 701, arXiv:hep-th/0305147. 
\bibitem{MarsSenovilla1993}
M.~Mars and J.~M.~M.~Senovilla, Axial symmetry and conformal Killing vectors, 
Class.~Quant.~Grav. {\bf 10} 1633 (1993), arXiv:gr-qc/0201045. 
\bibitem{Hiscock1981}
W.~A.~Hiscock, Models of evaporating black holes. I, 
Phys.~Rev.~{\bf D23} (1981) 2813. 
\bibitem{Wald_book1994}
R.~M.~Wald, Quantum Field Theory in Curved Spacetime and Black Hole Thermodynamics, 
The University of Chicago Press (1994). 
\bibitem{KellyWall2014}
W.~R.~Kelly and A.~C.~Wall, Holographic proof of the averaged null energy condition, 
Phys.~Rev.~{\bf D90} (2014) 106003. 
Erratum: [Phys.~Rev.~{\bf D91} (2015) 069902], arXiv:1408.3566 [gr-qc].
\bibitem{Rosso}
F.~Rosso, Global aspects of conformal symmetry and the ANEC in dS and AdS, 
JHEP 03 (2020) 186, arXiv:1912.08897 [hep-th]
\bibitem{Haddad2012}
N.~Haddad, Black Strings Ending on Horizons,  
Class.~Quant.~Grav. {\bf 29} (2012) 245001 arXiv:1207.2305 [hep-th]. 
\bibitem{IMM2017}
A.~Ishibashi, K.~Maeda, and E.~Mefford,  
Holographic stress-energy tensor near the Cauchy horizon inside a rotating black hole, 
Phys.~Rev.~{\bf D96} (2017) 024005 arXiv:1703.09743 [hep-th]. 
\end{thebibliography}
\end{document}